\documentclass[10pt,journal,compsoc]{IEEEtran}
\ifCLASSOPTIONcompsoc
  \usepackage[nocompress]{cite}
\else
  \usepackage{cite}
\fi

\ifCLASSINFOpdf
   \usepackage[pdftex]{graphicx}
   \usepackage{xcolor}
\else
\fi
\usepackage{url}

\usepackage{siunitx}

\usepackage{paralist} 

\usepackage[draft]{changes}
\definechangesauthor[name=Balbertini, color=blue]{BCA}
\usepackage{xcolor}

\newcommand{\urls}[1]{{\footnotesize \url{#1}}}
\newcommand{\msg}[1]{\textsc{#1}}
\newcommand{\virtioblk}{\texttt{virtio\_blk}}

\begin{document}
%
\title{Benchmarking the Security Protocol and Data Model (SPDM) for component authentication}
%
%
%


\author{Renan C. A. Alves,
        Bruno C. Albertini,
        Marcos A. Simplicio Jr
\IEEEcompsocitemizethanks{\IEEEcompsocthanksitem The authors are with Universidade de São Paulo, Brazil.}
}

\IEEEtitleabstractindextext{%

\begin{abstract}
%
%
Efforts to secure computing systems via software traditionally focus on the operating system and application levels.
In contrast, the Security Protocol and Data Model (SPDM) tackles firmware level security challenges, which are much harder (if at all possible) to detect with regular protection software.
SPDM includes key features like enabling peripheral authentication, authenticated hardware measurements retrieval, and secure session establishment.
Since SPDM is a relatively recent proposal, 
there is a lack of studies evaluating its performance impact on real-world applications.
In this article, we address this gap by: 
(1)~implementing the protocol on a simple virtual device, and then investigating the overhead introduced by each SDPM message; and 
(2)~creating an SPDM-capable virtual hard drive based on VirtIO, and comparing the resulting read/write performance with a regular, unsecured implementation.
Our results suggest that SPDM bootstrap time 
takes the order of tens of milliseconds, while the toll of introducing SPDM on hard drive communication highly depends on specific workload patterns. 
For example, for mixed random read/write operations, the slowdown is negligible in comparison to the baseline unsecured setup. 
Conversely, for sequential read or write operations, the data encryption process becomes the bottleneck, reducing the performance indicators by several orders of magnitude.
\end{abstract}

}
 %

\maketitle

\IEEEdisplaynontitleabstractindextext

%
\IEEEpeerreviewmaketitle


\IEEEraisesectionheading{\section{Introduction}\label{sec:introduction}}

\IEEEPARstart{M}{odern}
computing devices are commonly built using several components from different manufacturers.
This creates complex supply chains that usually end with an integrator, responsible for connecting off-the-shelf third-party components in an assembly line.
After assembly, devices are tested and then shipped to warehouses, eventually reaching the end user's facilities.
subsequently, the device components may undergo additional modifications, e.g., by means of firmware updates.
%

At every stage of the supply chain, there is a risk that malicious entities may inconspicuously tamper with or even substitute components \cite{firmware-attacks-ndss:2013,detecting-firmware-modification-sdd-current:2021}.
The goals of such attacks may vary.
For example, they may consist in eavesdropping sensitive data stored in volatile or non-volatile memory, as well as passing through video, audio or network components.
They may also involve modifying the default behavior of components to to enforce built-in obsolescence or impose some kind of censorship.

Although it may seem that such attacks at the hardware/firmware level belong to the realm of fiction, the existence of successful proof-of-concept implementations indicate otherwise.
For example, a vulnerability in the remote firmware update functionality of a family of printers allowed arbitrary code to be injected into the printer firmware \cite{firmware-attacks-ndss:2013}.
The authors then developed a self-replicating malware capable of eavesdropping on data and performing network reconnaissance.
Another attack, this time targeted at USB firmware, enabled a device (e.g., a thumb drive) to impersonate a different device type (e.g. a keyboard) \cite{firmware-validation-home-net:2016}.
Even more concerning is the fact that evidence of firmware tampering has also been found outside the halls of academia. 
In 2015, for example, hard drives from various manufacturers had their firmware altered in such a manner that the modified code could be used to retrieve data even from encrypted partitions \cite{nsadisk:2015}.

Protecting against such hardware-level tampering attacks is a challenging task.
In particular, the usual protection techniques that operate at the operating system level (e.g., antivirus software) or techniques that act at the infrastructure level (e.g., firewalls) are oblivious to such threats.
Aiming to overcome this issue, the Security Protocol and Data Model (SPDM) \cite{spdm-1.1.0:2020} was recently proposed by the DMTF (Distributed Management Task Force) to address such low-level security challenges. 
SPDM's main goals are to allow components to authenticate one another, to provide measurements of their internal state, and to securely exchange session keys.
Firmware measurements enable system components to be verified, ensuring they have not been victim of tampering, while establishing sessions keys avoids passive eavesdropping by malicious components attempting to steal data.

Albeit promising, SPDM is a relatively recent proposal, so its actual impacts on system performance have not yet been thoroughly evaluated in the literature.
This article aims to close this gap by measuring the overhead added by SPDM's security layer, assessing how it could impact the end-users' experience and identifying  bottlenecks that might be optimized in the protocol.
Specifically, we quantify the overhead added by the security layer to size the impact on end-user experience, and to identify protocol bottlenecks that might be optimized, using an emulated environment.
The resulting contributions are twofold:
(1) we evaluate the overhead of each individual phase of the SPDM protocol execution on an extremely simple SPDM-enabled component, a random number generator (Section~\ref{sec:uio}); and
(2) we build an SPDM-enabled hard drive and assess the performance impact on userspace-perceived reading and writing speeds (Section~\ref{sec:hd}).
All experiments build upon the virtualization capabilities of the QEMU emulator to implement our proof-of-concept devices and run the performance tests.
For easy reproducibility, and also because the SPDM-enabled artifacts developed as part of this work may be of independent interest, the corresponding source code is publicly available at \url{https://github.com/rcaalves/spdm-benchmark}.
In summary, our results show that SPDM certificate-based bootstrapping procedure takes around $66ms$.
Meanwhile, using SPDM to secure hard drive application data can greatly reduce the maximum transfer rate on sequential operations, but the impact was negligible on randomized mixed read/write workloads.

The rest of this manuscript is organized as follows.
Section~\ref{sec:spdm} summarizes SPDM's key aspects.
We study the individual overhead of each SPDM message in Section~\ref{sec:uio}, where we describe our methods and results.
Section~\ref{sec:hd} contains specification and results regarding our SPDM-enabled hard drive.
Finally, we discuss related work in Section~\ref{sec:related}, and present closing remarks in Section~\ref{sec:final}.
%

\section{SPDM} \label{sec:spdm}

This section summarizes basic concepts and workflow of SPDM in its version 1.1~\cite{spdm-1.1.0:2020}.

The SPDM is a proposed standard for secure intercommunication among hardware components.
It follows a requester-responder paradigm, and focuses on defining a set of useful operations and message formats that enable mutual authentication and the establishment of secure channels over an insecure medium.
At the same time, it aims to be agnostic to the physical medium and encapsulation approach employed for conveying those messages.

\begin{figure}[tb]
    \centering
    \includegraphics[width=0.9\columnwidth]{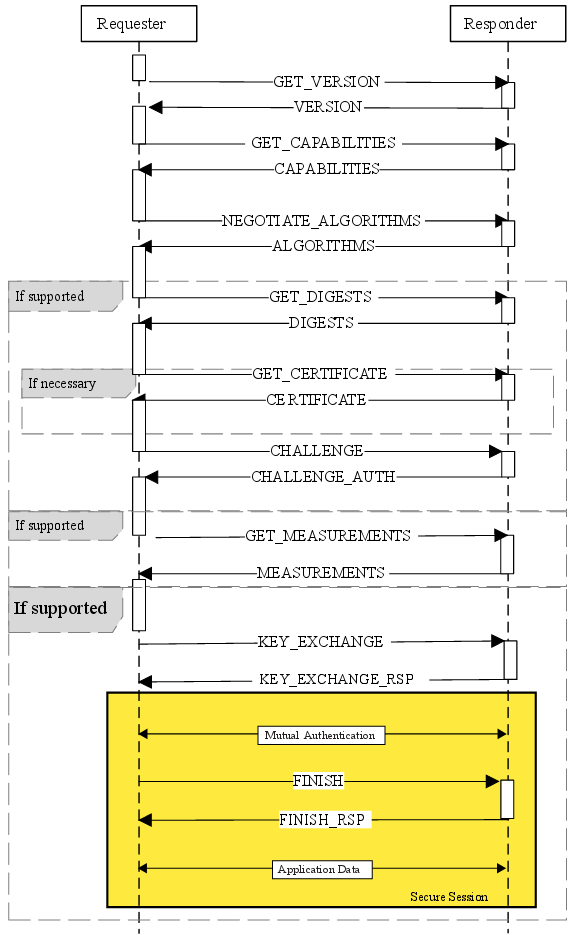}
    \caption{SPDM message flow. Extracted from \cite{spdm-1.1.0:2020}.}
    \label{fig:spdm_flow}
\end{figure}

The SPDM standard defines three pairs of mandatory messages, 14 pairs of facultative messages, and additional messages to handle errors.
A set of those messages are related to the core SPDM functions, such as device authentication, measurement retrieval, and secure session establishment.
The other messages serve the purpose of reporting available resources, stating which optional features are present, negotiating cryptographic algorithms to be employed, and maintaining an active communication session.
Figure~\ref{fig:spdm_flow} gives an overview of the expected message flow, highlighting which messages are mandatory.
Since each individual message is evaluated in our experiments, we briefly describe them in what follows.

The first pair of messages are \msg{get\_version} and \msg{version}. 
They are employed to settle on the SPDM version to be used. 
The protocol proceeds only if at least one of the versions advertised by the responder is supported by the requester.
Next, the requester inquires the responder about its capabilities, aiming to discover which optional messages are supported (messages \msg{get\_capabilities} and \msg{capabilities}).
The last pair of mandatory messages is \msg{negotiate\_algorithms} and \msg{algorithms}. 
They are exchanged so requester and responder can agree on the set of cryptographic algorithms they will use henceforth, throughout the protocol execution.
These mandatory messages are expected to present low overhead, since they have a slim payload and their processing does not involve any compute-intensive operation.

The next set of messages serves the purpose of retrieving certificates.
The \msg{get\_digest}/\msg{digest} message pair enables the requester to check whether any of the responder certificates has been previously fetched and cached. 
If that is not the case, certificates are retrieved via the \msg{get\_certificate}/\msg{certificate} message pair.
After obtaining the responder's certificate, the requester may challenge the responder to prove that it is the rightful owner of the corresponding public/private keys through the \msg{challenge}/\msg{challeng\_auth} message pair.
In this process, the responder may also indicate in its \msg{challeng\_auth} response that it would like to perform a mutual authentication, as long as this feature is supported by both responder and requester.
Mutual authentication follows the same procedure employed when authenticating the responder toward the requester (i.e., with the messages for fetching digests, requesting certificate, and issuing a challenge), but reversing the roles of each communicating component.
Since the responder typically does not send asynchronous messages to the requester, though, encapsulated messages are used instead: 
the requester sends a \msg{get\_encapsulated\_request} message to start the communication; 
the responder then answers with a \msg{encapsulated\_request} message, which triggers a \msg{deliver\_encapsulated\_response} reply by the requester; 
finally, the responder acknowledges the reception of this latter message by means of a \msg{encapsulated\_response\_ack} message, which may itself contain another encapsulated request if necessary.

A measurement may represent firmware, software, or some configuration data of an endpoint that helps to ensure that it is not counterfeit. 
The requester sends \msg{get\_measurements} message to request measurements, which is answered by a \msg{measurements} message.
The requester usually demands the responder to sign \msg{measurements} messages, although such a signature may also be omitted.

The requester can also issue a \msg{key\_exchange} message aiming to initiate a secure handshake for establishing a shared secret key.
The responder then answers with a \msg{key\_exchange\_rsp} message. 
The handshake is finalized by a \msg{finish}/\msg{finish\_rsp} message pair, when the secret key computed by both endpoints is confirmed and bound to the corresponding secure session established between them.
These messages rely on the responder being provisioned with at least one certificate chain considered valid by the requester.
Alternatively, if a pre-shared key is used instead of certificates, the messages above are replaced by their \msg{psk\_} counterparts, i.e., \msg{psk\_key\_exchange}, \msg{psk\_key\_exchange\_rsp}, \msg{psk\_finish}, and \msg{psk\_finish\_rsp}.

After a session is established, the underlying keys can be updated with the \msg{key\_update} and \msg{key\_update\_ack} messages. 
After a session is established, the underlying keys can be updated with the \msg{key\_update} and \msg{key\_update\_ack} messages; both requester and responder may initiate this key update process.
Also, if a session timeout period has been configured, \msg{heartbeat}/\msg{heartbeat\_ack} messages can be used to keep the session alive even in the absence of regular traffic.
When the session must be terminated, though, this can be accomplished by means of the \msg{end\_session}/\msg{end\_session\_ack} messages.

Finally, exceptions during the protocol execution can be flagged by \msg{error} messages, such as \msg{InvalidRequest}, \msg{Busy}, and \msg{DecryptError}.

\section{SPDM overhead assessment} 
\label{sec:uio}

The goal of this experiment is to assess the overhead introduced by each phase of the SPDM message flow.
To this end, we developed a random number generator (RNG) device that supports two operation modes: 1) secure mode, which follows the SPDM specification, and 2) clear-text mode, without any SPDM-related security capabilities.

\subsection{Method}
\label{sec:uio-method}

We used a virtualized experimental setup based on the QEMU emulation software \cite{qemu}.
The virtual machine run by QEMU contains an instance of our RNG device, which implements an SPDM responder.

The RNG was designed as a PCI device with a memory-mapped input/output (MMIO) region to send and receive SPDM messages, as well as to exchange control information.
After initializing, the RNG waits until a device driver writes a request message to the MMIO region, and then indicates that the message is ready by writing to a control register.
After reading the message, the RNG device sends an interrupt signal to announce that the response can be read from the MMIO region.

Non-SPDM transactions were used as a baseline for estimating the overhead introduced by SPDM to this simple procedure.
These transactions occur over specific memory region addresses, distinct from the region used for SPDM transactions.

\begin{figure}[htb]
    \centering
    \includegraphics[width=0.96\columnwidth]{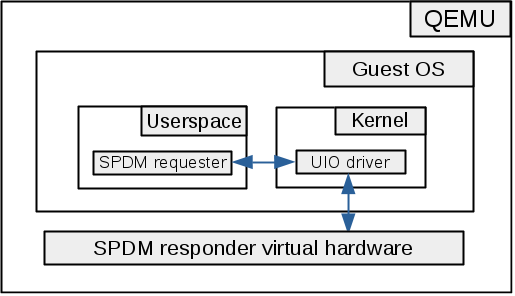}
    \caption{System architecture}
    \label{fig:rng_diagram}
\end{figure}

We implemented an SPDM requester as a device driver for our RNG on a Linux-based operating system. 
This device driver is built upon the userspace~I/O~system~(UIO).
Figure~\ref{fig:rng_diagram} depicts a schematic of our setup.
We note that, at least in principle, SPDM should be oblivious to the operating system. 
Nonetheless, the setup described provides a controlled and malleable environment, suitable for yielding meaningful performance results from our prototype implementation.

In our experiments, the interaction between the RNG device and its device driver was conducted following five steps:
1)~SPDM local initialization (memory allocation, library configuration);
2) SPDM connection, comprising version, capabilities, and algorithms negotiation;
3) SPDM authentication, including digest, certificate, and challenge messages;
4) measurement retrieval; and
5) application phase.

We analyzed the measurement messages in two ways: retrieving each measurement individually and retrieving all of them at once.
In the first case, the requester first inquires about the total number of measurements the responder holds before retrieving each one individually, while requiring a signature only for the last one.
Either way, the responder was configured to hold 5 different measurements, following the DMTF measurement block specification (SPDM specification v1.1.0~\cite{spdm-1.1.0:2020}, Section 10.11.1.1). 
Each block contains 128 bytes of dummy data.

The application phase is also divided in five steps: 
1) the endpoints agree on a shared key; 
2) a heartbeat message is transmitted; 
3) the session key is updated; 
4) the requester asks for a random number from the responder; and 
5) the session is concluded.
The session establishment process was tested both in the pre-shared key (PSK) and in the certificate-based settings.

We used the kernel's performance event infrastructure API to extract performance indicators~\cite{perf_book}, allowing us to assess each message individually.
We focused on two metrics provided by the \texttt{perf} API: number of cycles, and total CPU time.
CPU time provides a tangible sense of how long a task takes to finish. 
However, it depends on the underlying hardware and it can be deceiving in an emulated environment.
The number of cycles, on the other hand, is a more generic metric in comparison with CPU clock time measurements, although it remains platform dependent.
Furthermore, we configured \texttt{perf} to exclude from the count any event that occurs in kernel space, at the hypervisor, or at the guest machine (for responder only).
It is also worth mentioning that the \texttt{perf} API does not give perfect measurements, since the results provided include some overhead introduced by itself; this influence is, however, observed mostly on short measurements.

SPDM capability was provided by the \texttt{libspdm} open source library \cite{libspdm}.
For better reproducibility, \texttt{libspdm} compile-time and execution-time parameters used in the experiment are listed in Table~\ref{tab:openspdm_config}.

\begin{table}[htb]
\renewcommand{\arraystretch}{1.3}
    \centering
    \caption{libspdm parameters}
    \begin{tabular}{|c|c|}
        \hline
        Parameter & Value \\ \hline
        \hline
        SPDM protocol version & 1.1 \\ \hline
        \texttt{libspdm} version & commit \texttt{dc48779} \\ \hline
        \texttt{libspdm} build options & x64, release \\ \hline
        Underlying crypto library & MbedTLS \\ \hline
        Requester signature algorithm & TPM\_ALG\_RSAPSS\_3072 \\ \hline
        Responder signature algorithm & TPM\_ECC\_NIST\_P384 \\ \hline
        Measurement hash algorithm & TPM\_ALG\_SHA384 \\ \hline
        DHE algorithm & SECP\_384\_R1 \\ \hline
        AEAD algorithm & AES\_256\_GCM \\ \hline
        Key scheduling algorithm & As defined by SDPM v1.1 \\ \hline
        Mutual authentication & enabled \\ \hline
    \end{tabular}
    \label{tab:openspdm_config}
\end{table}

As emulation environment, we used QEMU version~4.1~\cite{qemu}. 
A QEMU virtual machine is specified by command-line options: hard drives, CPU, network cards, along with other options. 
The chosen guest operating system is based on \texttt{qemu\_x86\_64\_defconfig} preset configuration from Buildroot\footnote{Available at \url{https://buildroot.org/}} version~2020.02.9, which contains Linux Kernel version~4.19. 
Also for better reproducibility, the exact QEMU command-line options used in the experiment are listed below:

\begin{itemize}
    \item \texttt{-enable-kvm}: enables KVM (Kernel-based Virtual Machine), which enhances virtual machine performance. Needed so the guest kernel's performance event infrastructure is granted access to hardware counters;
    \item \texttt{-cpu qemu64,pmu=on}: selects emulated CPU model and enables Performance Monitoring Unit (PMU), needed to access performance counters;
    \item \texttt{-device spdm}: attaches an SPDM RNG device to the virtual machine;
    \item \texttt{-kernel bzImage}: selects the kernel booted by the virtual machine. We used Linux Kernel version 4.19 from Buildroot;
    \item \texttt{\footnotesize-drive file=rootfs.ext2,if=ide,format=raw}: indicates the root file system used in the VM, using IDE interface, and raw format. We used the root file system from Buildroot;
    \item \texttt{\footnotesize-append "console=ttyS0 rootwait root=/dev/sda"}: kernel command-line options. Sets the default console output, waits until root device is ready, and sets the root file system partition.
    \item \texttt{-m 1024}: sets the amount of RAM at the virtual machine, in megabytes
\end{itemize}

The host system ran a Linux-based system on an Intel \mbox{i7-10700KF} processor, with $3800 MHz$ clock, 8~Cores, 16~logical processors, and 32 GB of RAM.
Ideally, the CPU clock should be constant while performing benchmarks.
Hence, aiming to approach this ideal scenario, the following options were disabled at the machine BIOS configuration menu: turbo boost, speed step, speed shift, hyper threading, and CPU C states.

Each of execution step was performed 100 times, aiming to obtain statistical confidence. 
The graphs shown in Sections~\ref{sec:results_uio-requester} and \ref{sec:results_uio-responder} present the average value of all runs along with $95\%$ confidence intervals.

\subsection{Results: requester} \label{sec:results_uio-requester}

\begin{figure}[htb]
    \centering
    \includegraphics[width=0.95\columnwidth]{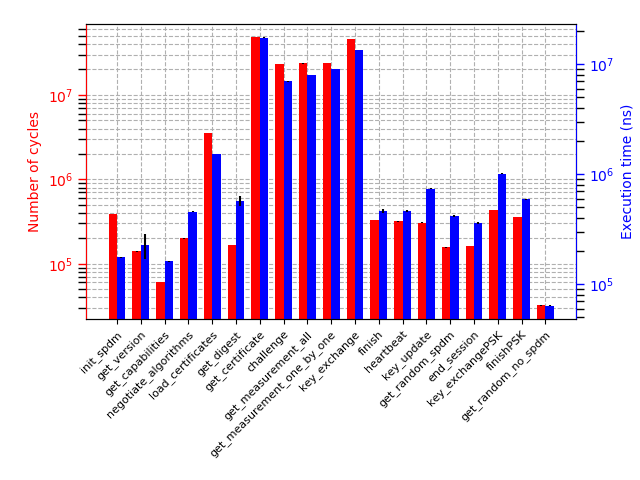}
    \caption{Requester execution time and number of cycles.}
    \label{fig:spdm_requester_overhead}
\end{figure}

Requester results are presented in Figure~\ref{fig:spdm_requester_overhead}, for both metrics hereby covered: cycle count and execution time.
As expected, most of the overhead was due to messages related to the authentication process.

Figure~\ref{fig:spdm_requester_overhead} shows that the most time consuming messages are \texttt{GetCertificate} and \texttt{KeyExchange}, taking respectively $17.4$ and $13.5 ms$, or $47.9$ and $45.9$ million cycles on average.
The \texttt{GetCertificate} procedure is expected to be slow since (1) it may require several messages to finish, and (2) by the end of it, the retrieved certificate must be verified for correctness, which requires a few signature verifications.
\texttt{KeyExchange}, in turn, involves the generation of a symmetric key pair by means of a Diffie-Hellman key exchange. 
Also, our analysis considers that the communication parties engage in mutual authentication, which can be considered a worst case scenario in terms of performance.

The next most time consuming messages are \texttt{GetMeasurements} and \texttt{Challenge}, both of which take around $23.5$ million of cycles to execute. 
The reason is that, in both cases, this overhead refers essentially to their underlying signature verification procedure, so it is reasonable for them to take roughly the same amount of time.
Interestingly, tough, we noticed that retrieving measurements all at once is slightly
faster than retrieving measurements one by one.
Specifically, the time taken in the former case is $7.9 ms$, against $9.0 ms$ in the latter. 
This represents a time gain of $11.9\%$, while the number of cycles is reduced~by~$2.8\%$.

The usage of Pre Shared Keys (PSK), on the other hand, considerably reduces the burden of establishing session keys.
The precise difference is that \texttt{KeyExchangePSK} takes only $1.0 ms$ (\num{4.3E+05} cycles), which is only a fraction of the $13.5 ms$ (\num{4.6E+07} cycles) observed in its asymmetric counterpart.
Also, using a PSK setting allows components to forego \texttt{GetCertificate} and \texttt{Challenge} messages, further speeding up the process.

All other messages take from $161$ up to $738 us$ on average, depending on their underlying complexity, so they are unlikely to become bottlenecks in practice.
Conversely, we notice that the task of loading the certificate from the disk can take a significant amount of time: $1.5 ms$, or \num{3.6E+06} cycles, on average.

Once a secure session is established, we were able to retrieve a random number from the SPDM-enabled RNG device in $415 us$ (\num{1.6E+05} cycles).
Conversely, in our SPDM-free baseline execution the same operation took $63 us$ (\num{3.2E04}~cycles) on average.
This means that SPDM led to a 6.4-fold increased in terms of time, or a 4.8-fold increase in number of cycles.
This result is not surprising, though, when we take into account the remarkable simplicity of the device evaluated in our tests.
After all, our simple RNG was designed to be extremely fast, so even cryptographic operations that are quite lightweight in absolute numbers, like symmetric encryption, become comparatively expensive.

\subsection{Results: responder} \label{sec:results_uio-responder}

\begin{figure}[htb]
    \centering
    \includegraphics[width=0.95\columnwidth]{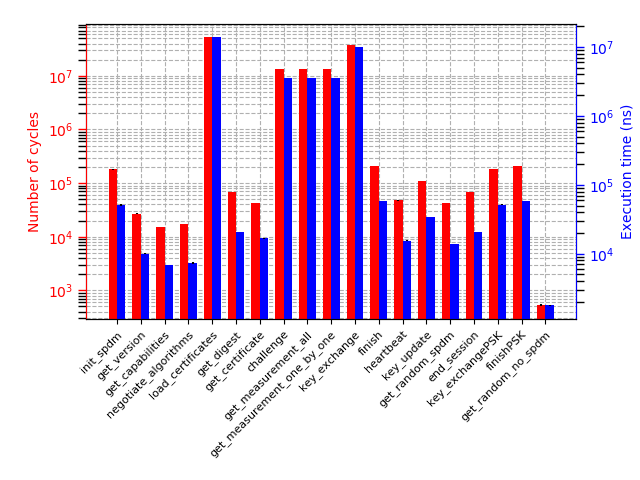}
    \caption{Responder execution time and number of cycles.}
    \label{fig:spdm_resonder_overhead}
\end{figure}

Figure~\ref{fig:spdm_resonder_overhead} shows metrics extracted from the Responder.
As shown in this figure, \texttt{KeyExchange} is the most expensive message to process, taking $10.1 ms$, or \num{3.8E+07} cycles.
As expected, though, adopting a PSK setting reduces the \texttt{KeyExchange} overhead to only $52 us$ or \num{1.8E+05} cycles.

In comparison with the requester, the responder handles \texttt{GetCertificate} messages faster than \texttt{KeyExchange}.
The reason behind this behavior is that most of the cryptographic processing of \texttt{GetCertificate} remains at the requester side. 
At the responder, \texttt{GetCertificate} takes $17 us$, or \num{4.2E+04} cycles.

Similarly to the requester, \texttt{GetMeasurements} and \texttt{Challenge} are nearly tied as the second most time consuming operations at the responder, taking approximately $13$ million cycles or $3.5 ms$.
Once again, retrieving measurements all at once was slightly faster than retrieving one by one.
More precisely, those operations take respectively $3.5$ and $3.6 ms$, which means a $1.5\%$ gain in terms of time costs (or a $1.1\%$ reduction in the number of cycles).

The largest overhead observed at the responder, however, was the time to load certificates from the disk, which took $13.8 ms$ or \num{5.2E+07} cycles. 
The discrepancy between responder and requester is caused by essentially by our configuration, where each party uses a different signature algorithm.
More precisely, the responder uses a signature algorithm based on elliptic curves, which takes longer to verify than the RSA-based signatures generated by the requester.

Processing a random number request without SPDM took the responder $1.8 us$, or $533$ cycles.
Adding the SDPM layer increases this cost to $13.8 us$, or \num{4.2E+04} cycles. 
The processing of all other messages took from $6.8 us$ to $58 us$ (or \num{1.5E+04} to \num{2.1E+05} cycles), adding a moderate amount of overhead to the protocol.

\section{Hard drive use case} \label{sec:hd}


This experiment was designed to assess the impact SPDM poses on system performance from a user perspective.
Among the possible peripherals typically found in computing systems, we chose to secure the communication between CPU and hard drive, both due to its importance and to the availability of tools for conducting such performance tests.
Specifically, we compared an SPDM-equipped hard drive to a regular, unsecured one, considering as metrics: boot time, read speeds, and write speed, under various workloads.

\subsection{Implementation details}


Due to the lack of off-the-shelf SPDM-enabled hardware, we once again resorted to emulation.
One more time, we used QEMU as emulation software, since it is equipped with a variety of open source virtual devices, including hard drives. 
After evaluating the available options, we chose to work with the \virtioblk\ hard drive.
The guest operating system, in turn, is a custom Buildroot-based Linux distribution. 
Its kernel contains a native \virtioblk\ driver, compatible with QEMU's \virtioblk\ hard disk.
Both driver and hard disk were then modified to incorporate SPDM security functionalities, as provided by \texttt{libspdm}.

\begin{figure}[htb]
    \centering
    \includegraphics[keepaspectratio=true, width=\columnwidth]{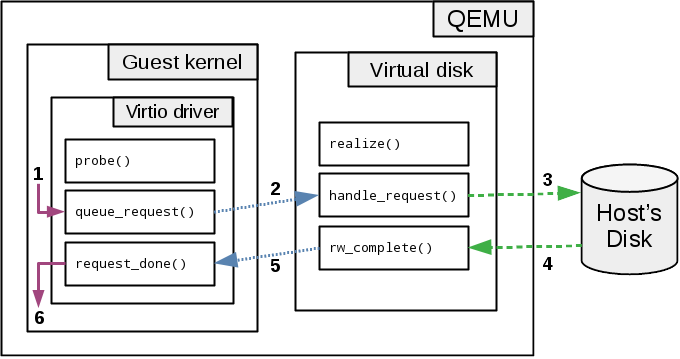}
    \caption{\virtioblk\ driver and virtual hardware interaction: 1) Guest operating system request, 2) Driver request to virtual hardware, 3) Request forwarded to host, 4) Request result at virtual hardware, 5) Request result at driver, 6) Request finished}
    \label{fig:virtioblk_diagram}
\end{figure}

In short, the interaction between kernel driver and virtual device is as follows: 
1)~the operating systems sends read/write requests\footnote{There are other kinds of operations, but we focus on reading and writing for illustrative purposes} to the driver queue (\texttt{queue\_request} function);
2)~the request is executed and
transferred from the guest virtual machine's kernel space to the virtual disk's request handler, triggering the \texttt{handle\_request} function;
3)~the request is forwarded to the host's disk;
4)~QEMU receives the request results from the host OS (\texttt{rw\_complete} callback function);
5)~QEMU forwards the results to the \virtioblk\ driver, activating the \texttt{request\_done} function;
6)~the guest kernel is informed that the I/O operation is complete.
The diagram in Figure~\ref{fig:virtioblk_diagram} illustrates these steps.

The aforementioned interaction was adapted to follow the SPDM workflow.
In the adaptation, the driver fills the role of the SPDM requester, while the hard drive takes the role of SPDM responder.
Requester and responder are required to bootstrap the SPDM protocol before heading into the application phase, in which read/write requests are encrypted.

The \virtioblk\ hard drive initializes internal SPDM variables and loads certificates during the virtual machine initialization (in the \texttt{realize} function). 
After calling this function, the hard drive is ready to process incoming SPDM messages.

The kernel driver performs a similar initialization when a new \virtioblk\ device is detected, using the \texttt{probe} function. 
At that point, not only are local variables initialized, but the whole SPDM bootstrap procedure also takes place. 
As a result, all SPDM messages, from \textsc{get\_version} to \textsc{key\_exchange}, are exchanged at this moment.
All SPDM messages are encoded similarly to regular read/write requests, but using a special operation code.
By the end of \texttt{probe} function, driver and hard disk obtain a symmetric key they can use for encrypting application data.

Incoming write requests are now encrypted as part of the \texttt{queue\_request} function in the kernel driver, and decrypted at the \texttt{handle\_request} function when it reaches the virtual disk. 
Analogously, a read request is encrypted after the data is retrieved from the host by means of the \texttt{rw\_complete} function (on QEMU), and decrypted by the kernel driver as part of the \texttt{request\_done} function.

Within both kernel and QEMU, SPDM functionalities are provided by \texttt{libspdm} using the same parameters employed in Section \ref{sec:uio-method}, as summarized in Table~\ref{tab:openspdm_config}.
Minor adjustments were needed to deal with stack overflow issues when calling some \texttt{libspdm} functions, though, caused by the limited stack space available within the Linux kernel.

\subsection{Method}

In our experiments, we used a few widely employed tools and benchmarking utilities to assess hard drive performance: \texttt{dd}, \texttt{hdparm}, \texttt{ioping}, \texttt{bonnie++}, \texttt{fio}, which are further discussed separately at the end of this section.
All tests were executed on a QEMU virtual machine, on a separately attached hard drive using the \virtioblk\ interface. 
Once again, the following CPU attributes were disabled at the machine BIOS configuration menu: turbo boost, speed step, speed shift, hyper threading, and CPU C states.
For each tool, two batches of experiments were executed: 
1) a baseline setting, with the unmodified device and driver; and 
2) an SPDM-enabled setting, with our modified implementations. 
In both scenarios, the following QEMU command line parameters were used:

\begin{itemize}
    \item \texttt{-enable-kvm}: enables KVM (Kernel-based Virtual Machine), which enhances virtual machine performance;
    \item \texttt{-cpu qemu64}: for selecting the emulated CPU model;
    \item \texttt{-kernel bzImage}: selects the kernel booted by the virtual machine. We used Linux Kernel version 4.19 from Buildroot;
    \item \texttt{\footnotesize-drive file=rootfs.ext2,if=ide,format=raw}: indicates the root file system used in the VM, using IDE interface, and raw format. We used the root file system from Buildroot;
    \item \texttt{\footnotesize-append "console=ttyS0 rootwait root=/dev/sda"}: kernel command-line options. Sets the default console output, waits until root device is ready, and sets the root file system partition;
    \item \texttt{-m 1024}: virtual machine RAM, in megabytes;
    \item \texttt{\footnotesize-drive file=benchmarkdisk,if=virtio,format=raw}: appends a virtio-based additional hard drive, which is the target of our experiments.
\end{itemize}

All experiments were conducted to reach statistical significance. 
Some of the tools used provide statistical data, while others that do not were run multiple times and had their outcomes summarized manually by the research team to achieve this goal.
The usage, metrics provided, and output processing for the five tools employed are described in what follows. 
%

\subsubsection{\texttt{\emph{dd}} (from BusyBox v1.31.1)}

this is a commonplace utility found on Unix-like operating systems.
Its primary usage is to transfer raw data from one destination to another.

In our experiments, we used \texttt{dd} to test write speed. 
Specifically, we read data from \verb|/dev/zero|, which is a fast source of dummy data, and wrote it to a file on the target disk.

We tested writing 2 gigabytes of data to the disk according to two approaches: in small blocks of 4KB each, or in 512MB-long blocks. 
In all cases, we enforced that the data was physically written on the device before the commands returned with the \verb|conv=fsync| command line option.
The write speed is calculated by the quotient between the total amount of data written and the time it takes to complete the operation. 
We used the \verb.time. command to measure the execution time with \verb.dd..
For each block size, the results hereby presented correspond to the average for 10 repetitions of the writing procedure.

\subsubsection{\texttt{\emph{hdparm}} (from BusyBox v1.31.1)}

this command line tool is also commonly found in Linux systems. 
Besides using it to set and read hard drive parameters, we also explores its \verb.-t. option switch, which provides buffered reading speed estimates.
The tests with this tool were run a total of 10 times.

\subsubsection{\texttt{\emph{ioping}} v0.9}

this is a tool for monitoring disk latency.
It works similarly to the well known tool from the network domain \verb.ping., i.e., by sending short requests to the disk and measuring how long they take to be fulfilled. 
We used the default parameters while testing both reading and writing latency, executing a total of 10 pings for each operation.

\subsubsection{\texttt{\emph{bonnie++}} v1.04}

this is a purpose-built benchmarking toolkit for hard drives. 
It automatically performs write, rewrite, and read tests. 
The metrics extracted from this tool were read and write speed measured in kB/s. 
The main command line options employed were:
\begin{itemize}
    \item \verb.-x 10. runs the benchmark 10 times;
    \item \verb.-s 2G. specifies total amount of read/write data to 2 gigabytes;
    \item \verb.-n 0. disables file creation test, which is of little interest to our scenario because since this relates mostly to the file system;
    \item \verb.-f. skips per-character tests, since our goal was to test HD behavior that is close to common system usage;
    \item \verb.-b. specifies unbuffered writes,
    \item \verb.-D. uses the \verb.O_DIRECT. flag, which attempts to perform requests synchronously.
\end{itemize}

\subsubsection{\texttt{\emph{fio}} v3.23}

this is a highly customizable benchmarking tool for hard drives. 
Its main goal is to enable the creation of a workload as close as possible to the desired test case. 
Among the large set of metrics provided by this tool, we focused on I/O operations per second (iops).
The main command line options used were:

\begin{itemize}
    \item \verb.--size=<size>. sets the portion of the disk that will be used to perform the tests. We used 2 GB in all experiments,
    \item \verb.--io_size=<size>. total amount of data used in each I/O transaction. We used 5 GB in all experiments,
    \item \verb.--rw=<option>. the type of I/O pattern. Common values are \verb.read. (sequential reads), \verb.write. (sequential writes), \verb.randread. (random reads), and \verb.randrw. (random reads and writes mixed),
    \item \verb.--blocksize=<size>. the size of each individual operation. We used 1024 KB for sequential operation patterns and 4 KB for random operation patterns,
    \item \verb.--fsync=<n>. issues a synchronization command at every \verb.<n>. writes.
    We configured \verb.<n>. as 10,000 for sequential operation patterns and 1 for random operation patterns.
\end{itemize}

The aforementioned set of command line options yields four different tests performed with the \texttt{fio} tool: 
1) sequential reads with 1024 KB blocks;
2) sequential writes with 1024 KB blocks; 
3) random reads with 4 KB blocks; and 
4) mixed reads and writes with 4 KB blocks.

\subsubsection{boot time}

contrary to the other metrics hereby evaluated, we did not use any specialized tool to measure the system boot time. 
Instead, we modified the guest's initialization scripts to log the system uptime as the last step of the initializing process. 
The system uptime was obtained from reading \verb./proc/uptime., a file that counts the seconds elapsed from the moment the kernel takes control of the CPU, yielding a precision of hundredths of a second. 
We collected a total of 15 boot times.

\subsection{Experimental Results}

This section presents and discusses the results obtained from each of the benchmark tools used. 

\subsubsection{\texttt{dd}}


\begin{figure}[htb]
    \centering
    \includegraphics[width=0.85\columnwidth]{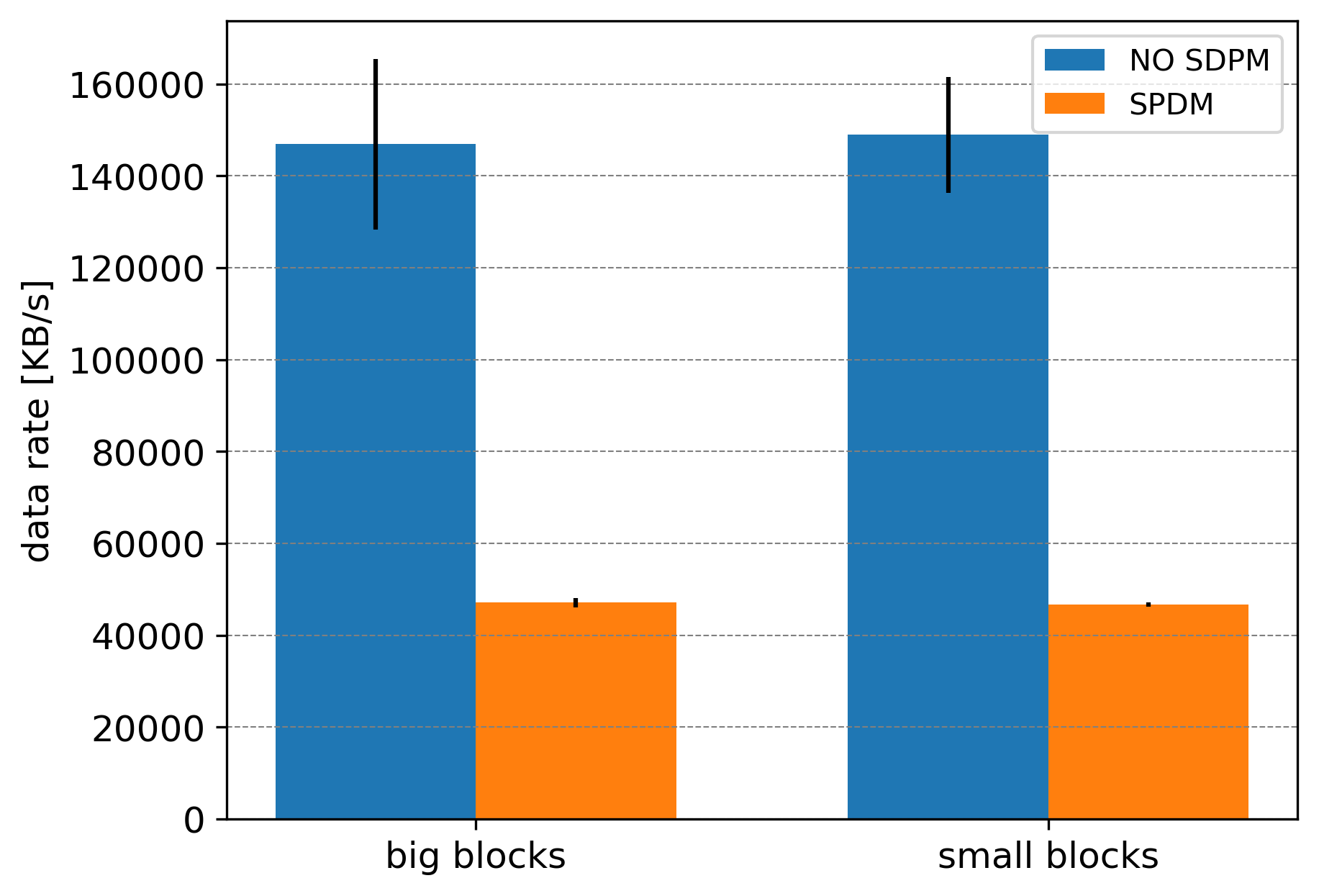}
    \caption{Measuring data rate of an SPDM-enabled hard disk: \texttt{dd}}
    \label{fig:dd}
\end{figure}

Figure~\ref{fig:dd} shows the results obtained from the \verb.dd. command. 
This experiment assesses the speed of writing data sequentially, which reduces the number of seek operations executed during the test. 
Hence, besides providing insights on SPDM's impact over disk writing speed when such workloads are prevalent, it also serves as baseline for scenarios where random disk accesses are more common.

In general, writing small blocks tends to be slightly slower than writing large blocks, which was observed in our results ($\approx$1\% slower). 
In both cases, though, SPDM caused a $\approx$68\% slowdown in writing speed.

\subsubsection{\texttt{hdparm}}


Quoting its manual, the hdparm test provides \emph{``an indication of how fast the drive can sustain sequential data reads under Linux, without any filesystem overhead.''}.
Without SPDM, the average read speed observed was $3.9 GB/s$, fairly close to the nominal $6 GB/s$ speed of the hard drive, considering the virtualization overhead.
Introducing SPDM, though, drastically decreases the value indicated by hdparm to $28 kB/s$, which translates to a $99.3\%$ speed degradation.

\subsubsection{\texttt{ioping}}

\begin{figure}[htb]
    \centering
    \includegraphics[width=0.85\columnwidth]{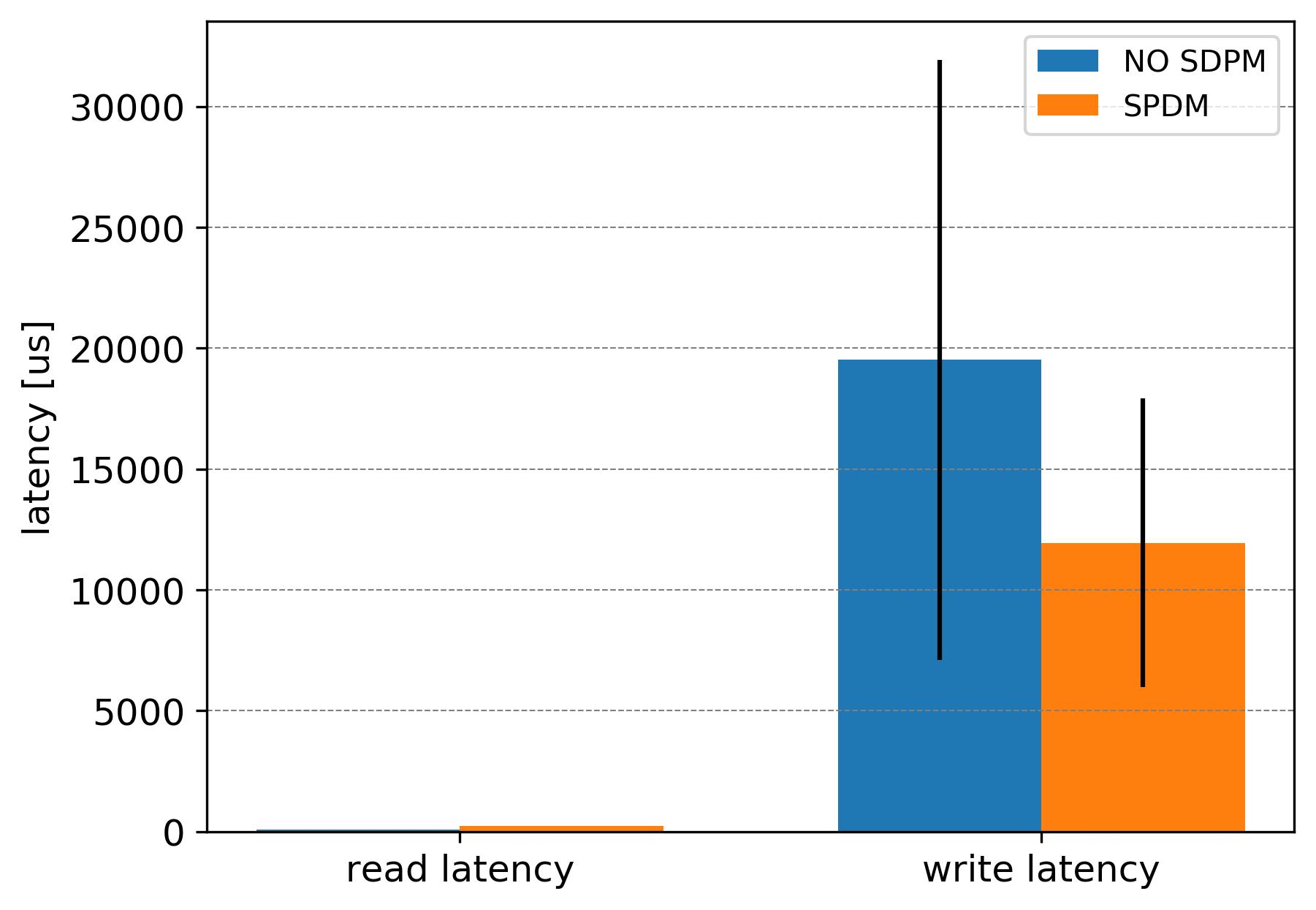}
    \caption{Measuring latency of an SPDM-enabled hard disk: \texttt{ioping}}
    \label{fig:ioping}
\end{figure}

Figure~\ref{fig:ioping} shows read and write latency results according to \verb.ioping..
Focusing on average results only, the introduction of SPDM increased the reading latency by $208\%$, while it decreased writing latency  by $39\%$. 
However, the obtained confidence intervals in both cases were very large, making it hard to draw statistically relevant conclusions with this tool.

\subsubsection{\texttt{bonnie++}}

\begin{figure}[htb]
    \centering
    \includegraphics[width=0.85\columnwidth]{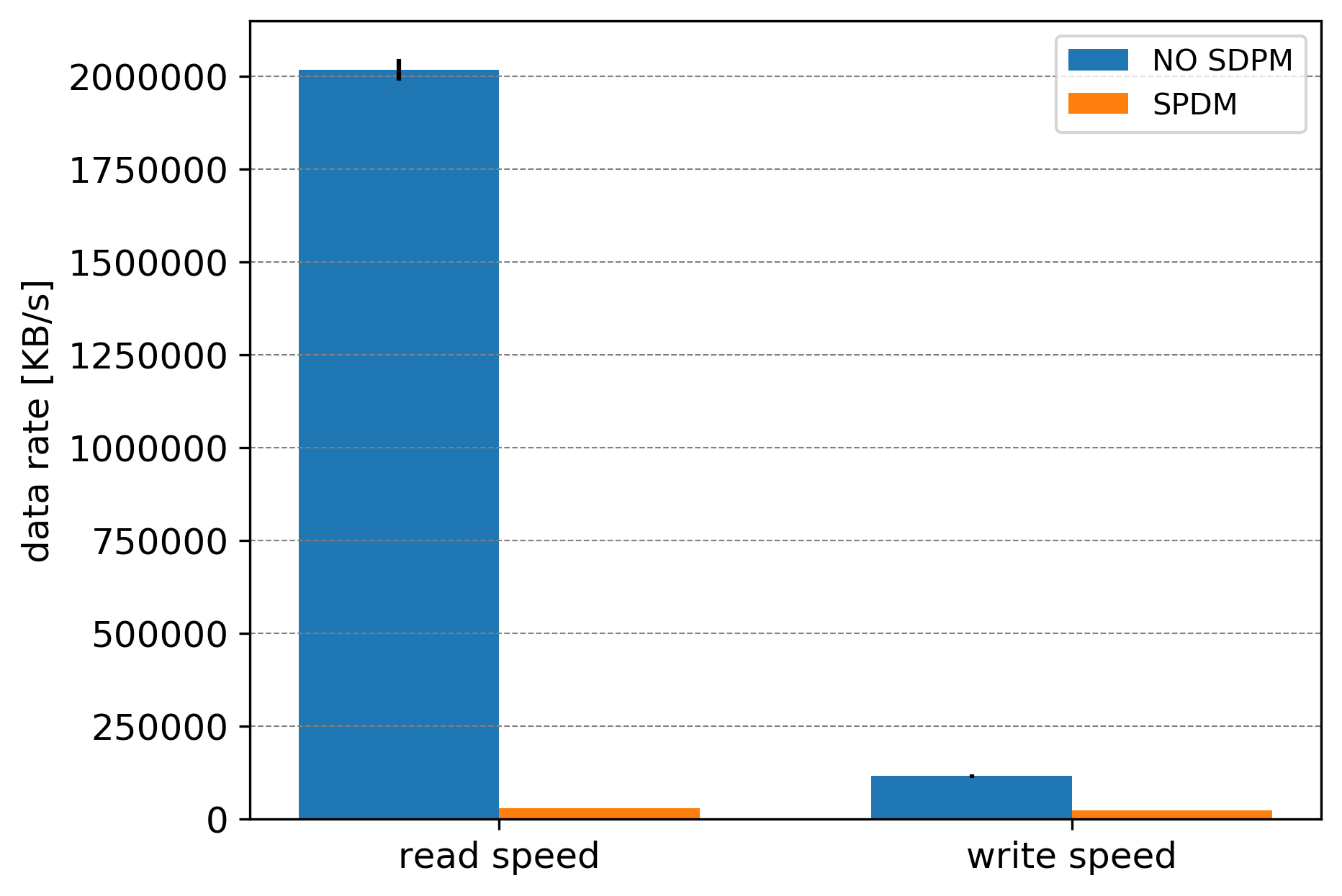}
    \caption{Measuring data rate of an SPDM-enabled hard disk: \texttt{Bonnie++}}
    \label{fig:bonnie}
\end{figure}

The results from \texttt{bonnie++} (Figure~\ref{fig:bonnie}) indicate a harsh loss of performance when SPDM is introduced.
The tests performed by this tool consist of reading and writing $200Mb$ files to the disk. 
The writing portion is somewhat similar to the one performed by the \verb.dd. writing test, but spread across multiple files.
However, the numbers show a deeper performance chasm than what was observed with  \verb.dd.: the writing speed drops from $115MB/s$ to $23MB/s$ (a $79.8\%$ reduction).
The loss of reading performance is even more significant: from $2.0GB/s$ to only $28MB/s$.
It makes sense that both reading and writing speed drop to the same order of magnitude, since the system bottleneck is the same in both tests -- the processing cost of encrypting/decrypting every transaction.

\subsubsection{\texttt{fio}}

\begin{figure}[htb]
    \centering
    \includegraphics[width=0.85\columnwidth]{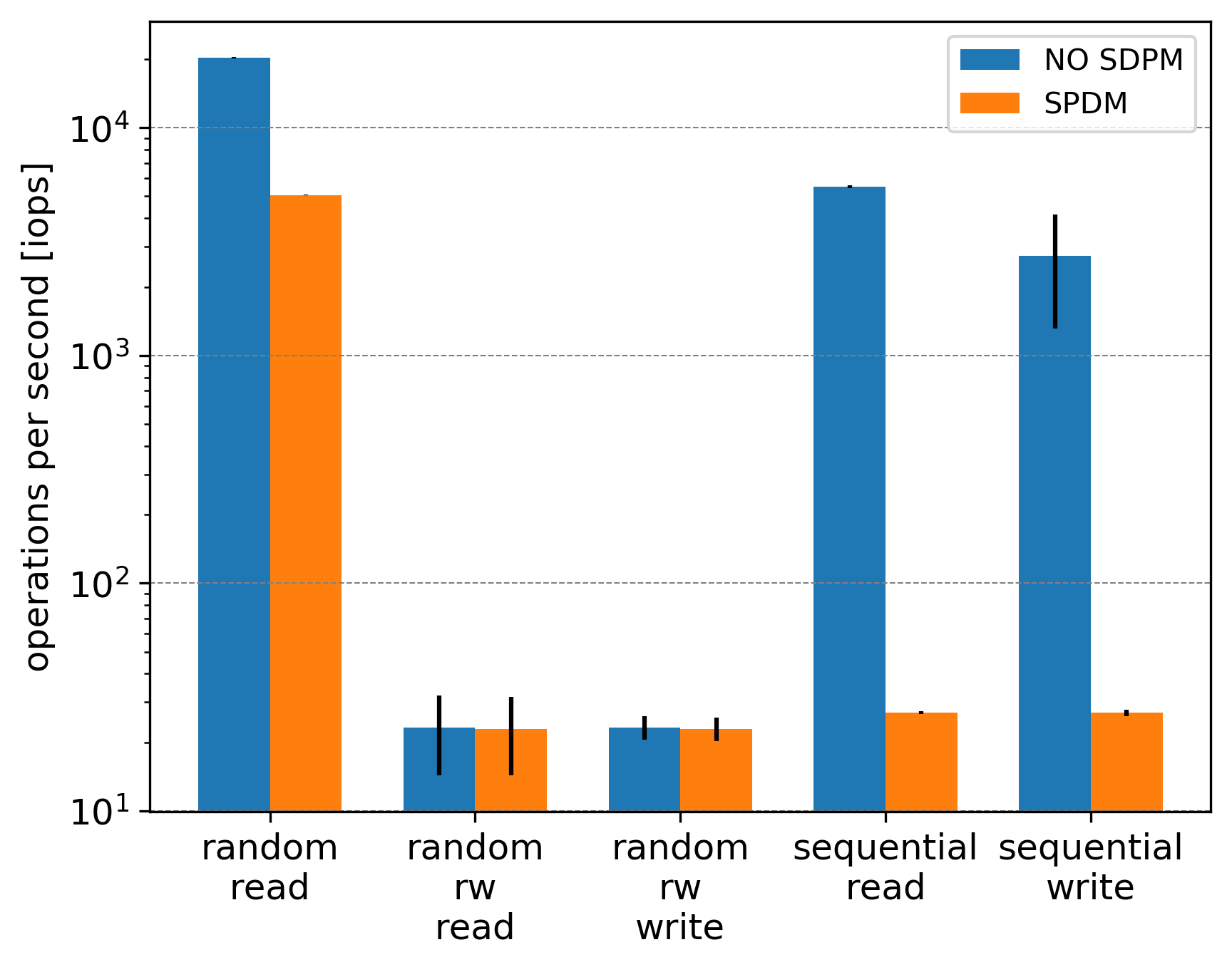}
    \caption{Measuring transaction rate of an SPDM-enabled hard disk: \texttt{fio}}
    \label{fig:fio}
\end{figure}

Results from the \texttt{fio} tool are presented in Figure~\ref{fig:fio}, where the unit of measure is input/output operations per second (iops).
The sequential tests (i.e., those labeled "sequential read" and "sequential write") consider large blocks while requesting synchronization sparsely. 
In this case, the pattern observed is similar to the one obtained with \texttt{dd} and \texttt{bonnie++}: there is a significant loss of performance, with the number of iops in the SPDM-enabled disk being less than $1\%$ of the baseline value.

The performance degradation becomes less prominent when randomness is introduced.
For these tests, the block size was 4kB, and requests to synchronize data were sent after every operation.
When performing random read-only tests, the performance drops to $25.0\%$ of the baseline value.
Mixing random reading and random writing (in Figure \ref{fig:fio}, label "random rw read" refers to read speed, and label "random rw write", refers to read speed) yields virtually the same level of iops: introducing SPDM causes a reduction of approximately $1.3\%$ in both cases, but the standard deviation width prevents us from attesting statistical difference.
We conjecture that the reason behind this trend is the bottleneck shifting from the cryptographic operations to the physical disk operations, namely the frequent seek operations to address the random request locations.

\subsubsection{Boot time}


During OS initialization, the SPDM-enabled HD driver performs all SPDM bootstrapping procedures, including loading certificates and establishing a symmetric key. 
Our test shows that these procedures increase OS boot time by $66ms$ ($3.48s$ to $3.55 s$), which resonates with the results presented in Section~\ref{sec:uio}.

\subsubsection{Summary}

Table~\ref{tab:hd_results} summarizes all numeric values discussed in this section.

\begin{table}[htb]
\caption{Hard drive benchmarks numeric results }
\scriptsize
\label{tab:hd_results}
\centering
\begin{tabular}{l|r|r|r|r|}
\cline{2-5}
                                                                                                & \multicolumn{1}{c|}{\begin{tabular}[c]{@{}c@{}}Average\\ W/ SPDM\end{tabular}} & \multicolumn{1}{c|}{\begin{tabular}[c]{@{}c@{}}Standard\\ Deviation\end{tabular}} & \multicolumn{1}{c|}{\begin{tabular}[c]{@{}c@{}}Average\\ w/o SPDM\end{tabular}} & \multicolumn{1}{c|}{\begin{tabular}[c]{@{}c@{}}Standard\\ Deviation\end{tabular}} \\ \hline
\multicolumn{1}{|l|}{\begin{tabular}[c]{@{}l@{}}dd small\\ blocks {[}kB/s{]}\end{tabular}}      & 4.66E+04                                                                       & 466                                                                               & 1.49E+05                                                                        & 1.27E+04                                                                          \\ \hline
\multicolumn{1}{|l|}{\begin{tabular}[c]{@{}l@{}}dd big\\ blocks {[}kB/s{]}\end{tabular}}        & 4.71E+04                                                                       & 1.01E+03                                                                          & 1.47E+05                                                                        & 1.86E+04                                                                          \\ \hline
\multicolumn{1}{|l|}{\begin{tabular}[c]{@{}l@{}}ioping read\\ latency {[}us{]}\end{tabular}}    & 234                                                                            & 12.0                                                                              & 76.0                                                                            & 4.00                                                                              \\ \hline
\multicolumn{1}{|l|}{\begin{tabular}[c]{@{}l@{}}ioping write\\ latency {[}us{]}\end{tabular}}   & 1.19E+04                                                                       & 5.98E+03                                                                          & 1.95E+04                                                                        & 1.24E+04                                                                          \\ \hline
\multicolumn{1}{|l|}{\begin{tabular}[c]{@{}l@{}}hdparm read\\ speed {[}kB/s{]}\end{tabular}}    & 2.81E+04                                                                       & 126                                                                               & 3.93E+06                                                                        & 7.29E+05                                                                          \\ \hline
\multicolumn{1}{|l|}{\begin{tabular}[c]{@{}l@{}}bonnie read\\ speed {[}kB/s{]}\end{tabular}}    & 2.83E+04                                                                       & 131                                                                               & 2.02E+06                                                                        & 2.89E+04                                                                          \\ \hline
\multicolumn{1}{|l|}{\begin{tabular}[c]{@{}l@{}}bonnie write\\ speed {[}kB/s{]}\end{tabular}}   & 2.32E+04                                                                       & 30.1                                                                              & 1.15E+05                                                                        & 4.22E+03                                                                          \\ \hline
\multicolumn{1}{|l|}{\begin{tabular}[c]{@{}l@{}}fio sequential\\ read {[}iops{]}\end{tabular}}  & 27.0                                                                           & 0.315                                                                             & 5.52E+03                                                                        & 86.2                                                                              \\ \hline
\multicolumn{1}{|l|}{\begin{tabular}[c]{@{}l@{}}fio sequential\\ write {[}iops{]}\end{tabular}} & 26.9                                                                           & 0.892                                                                             & 2.74E+03                                                                        & 1.43E+03                                                                          \\ \hline
\multicolumn{1}{|l|}{\begin{tabular}[c]{@{}l@{}}fio random\\ read {[}iops{]}\end{tabular}}      & 5.06E+03                                                                       & 15.6                                                                              & 2.02E+04                                                                        & 1.60E+02                                                                          \\ \hline
\multicolumn{1}{|l|}{\begin{tabular}[c]{@{}l@{}}fio random\\ rw read {[}iops{]}\end{tabular}}   & 22.9                                                                           & 8.63                                                                              & 23.2                                                                            & 8.90                                                                              \\ \hline
\multicolumn{1}{|l|}{\begin{tabular}[c]{@{}l@{}}fio random\\ rw write {[}iops{]}\end{tabular}}  & 23.0                                                                           & 2.70                                                                              & 23.3                                                                            & 2.75                                                                              \\ \hline
\multicolumn{1}{|l|}{boot time {[}s{]}}                                                         & 3.55                                                                           & 1.04                                                                              & 3.48                                                                            & 1.04                                                                              \\ \hline
\end{tabular}
\end{table}

\section{Related Work} \label{sec:related}

SPDM is a fairly new standard, so its impact and benefits have not been thoroughly evaluated in the literature.
Nevertheless, since SPDM's goal is to secure a system from early boot until OS runtime, it relates to other approaches that focus on pre-OS stages.
In particular, SPDM is somewhat close to techniques often called ``secure boot'' or ``trusted boot'', whose goal is to ensure that only legitimate initialization firmware and bootloaders are executed.

The literature includes a number of studies on secure boot performance.
For example, Profentzas et. al~\cite{8804799}
evaluate the overhead of software-based and hardware-based secure boot on embedded platforms, namely, raspberry pi and beaglebone. 
Their study shows that the secured system presents an increased boot time ranging from $4\%$ to $36\%$, depending on the technique and algorithms employed.

In a similar fashion, Khalid et. al~\cite{6581569} proposed a trusted boot architecture for embedded systems based on an independent security processor.
Their design was implemented on FPGAs, and their experiments showed that their secured boot process increases boot time by $25\%$, considering a Linux system customized for their needs.

The study by Yin et. al~\cite{6120953} brought attention to failure-prone NAND flash chips, commonly used to store bootloaders in embedded platform. 
The authors propose a redundancy scheme that verifies the integrity of bootloader firmware code and falls back to an alternative copy in case of checksum mismatch.
Their experiments show that total boot time is increased by $65\%$ if the bootloader is intact.
Otherwise, it is increased by $255\%$, since the backup code has to be copied and verified once again.

Kumar et. al~\cite{9116252} implemented a secure boot design based on post-quantum cryptography (PQC). 
Their main concern is that PQC algorithms require more computing resources than classic algorithms, which led them to create a custom FPGA implementation. 
Their experiments show that their implementation of the chosen PQC algorithm is at least 10 and at most 30 times slower than the baseline elliptic curve algorithm, depending on the level of parallelism.

Contrasting with the previous studies, Dasari and Madipagda~\cite{9255660} stands out for being one of the few works in the literature that include an analysis of SPDM.
Specifically, the authors propose an architecture to detect component tampering in end products, producing a birth certificate comprising the platform as a whole. 
They use SPDM as part of their solution to retrieve firmware hashes (measurements) from the end product individual components. 
However, their solution is based on a older version of SPDM, which did not yet support session key establishment.
Consequently, the solution does not prevent passive sniffers from eavesdropping sensitive information exchanged among components.
At the same time, and contrary to this work, there is no evaluation of the impact brought by SPDM on application level performance.

When considering the particular scenario of protecting communications from/to hard disks, this works shares a relationship with studies covering disk encryption technologies.
Examples include works that focus on the energy consumption of full disk encryption technologies \cite{disk-encryption-energy:2012}, on the impact of different cryptographic algorithms \cite{disk-encryption-algorithms-bench:2018}, or on specific types of devices \cite{disk-encryption-flash-devs:2011,disk-encryption-Samsung-ssd:2010,disk-encryption-seagate:2010}.
There are also more holistic studies, like \cite{sec.1028}, where architectures including secure boot and hard drive encryption mechanisms are proposed (in this particular case, for mobile devices).
Like the present study, such works give insights on how encryption impacts communications with hard disks, covering similar metrics.
The similarity ends there, though.
After all, SPDM's secure sessions are meant to protect in-transit messages, not only the actual data written to or read from the disk. 
Therefore, SPDM cannot be used as an alternative to disk encryption, since all encrypted data sent to the disk is decrypted upon reception.
Also, if disk encryption tools are employed together with SPDM, the protocol remains oblivious to the fact that it is protecting payloads that are already in encrypted form.

All in all, we note that the aforementioned studies do not tackle the impact of SPDM, and its corresponding runtime security between components, at the application level. 
Therefore, their results are not directly comparable to the experiments presented herein. 
To the best of our knowledge, this is the first work to assess the impact of SPDM's security overhead on userspace metrics.

\section{Final remarks} \label{sec:final}

The Security Protocol and Data Model (SPDM) aims at providing standardized ways for component (mutual) authentication, firmware integrity check, and secure communication establishment.

Although these functionalities are important to increase the security level of modern computing systems, it is expected to bring performance penalties.
Our goal in this paper is to assess the magnitude of SPDM's performance impact.
To the best of our knowledge, this is the first study that takes on this endeavor.

In summary, our results show that the overhead introduced by the most time-consuming SPDM message is $17.4ms$~(47.9~million~cycles), while the fastest messages take only a few microseconds. 
According to our experiments, the typical SPDM bootstrap takes approximately $50ms$ to run.
Regarding the hard drive benchmarks, we noticed that the specific workload greatly influences the final outcomes.On sequential read or write operations, data encryption becomes the bottleneck, and heavily affects performance.

On sequential read or write operations, data encryption becomes the bottleneck, and heavily affects performance (e.g., reading speed dropped from $2.0GB/s$ to $28MB/s$ in one run of the benchmark).
That is not the case, though,
for workloads
that are mainly comprised of random read and write operations scattered throughout the disk. 
For such workloads, we found no significant performance differences between the secured system and the baseline system, since the bottleneck is the physical movement of disk heads and switching between reading and writing modes.

As future work, we intend to explore further the performance impacts of deploying SPDM in modern systems.
This includes integrating SPDM with other classes of commonplace devices, such as network cards, and evaluating the protocol's overhead at earlier stages of the boot process.


\section*{Acknowledgment}

The authors would like to thank Hewlett Packard Enterprise for supporting this project.
FAPESP 2020/09850-0.

\ifCLASSOPTIONcaptionsoff
  \newpage
\fi

\bibliographystyle{IEEEtran}
\bibliography{IEEEabrv,refs}

\begin{IEEEbiography}
[{\includegraphics[width=1in,height=1.25in,clip,trim={1mm 2mm 46mm 0mm}]{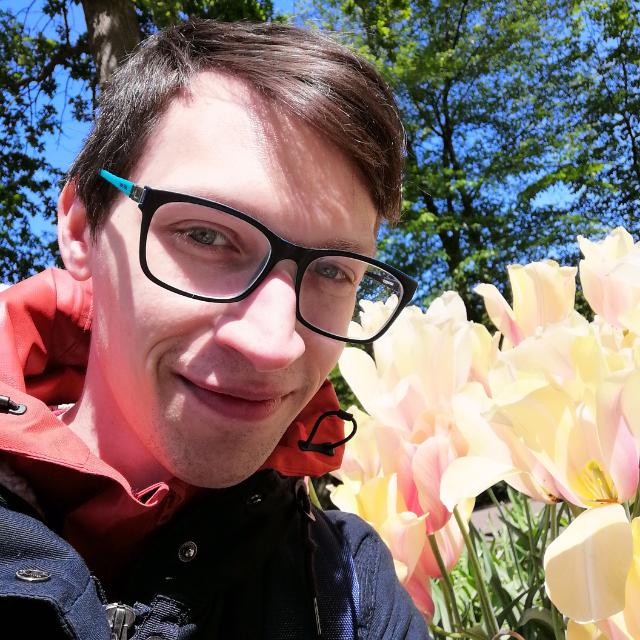}}]
{Renan C. A. Alves}
received the B.Sc. degree in
electrical engineering with emphasis on computer
and digital systems, the M.Sc. degree, and PhD from the
Universidade de São Paulo, Brazil, in 2011 and
2014, and 2020, respectively, where he is currently a post-doc and part-time professor. His main research interests
include protocol modeling, performance analysis, internet of things, and cybersecurity.
\end{IEEEbiography}

\begin{IEEEbiography}[{\includegraphics[width=1in,height=1.25in,clip,keepaspectratio]{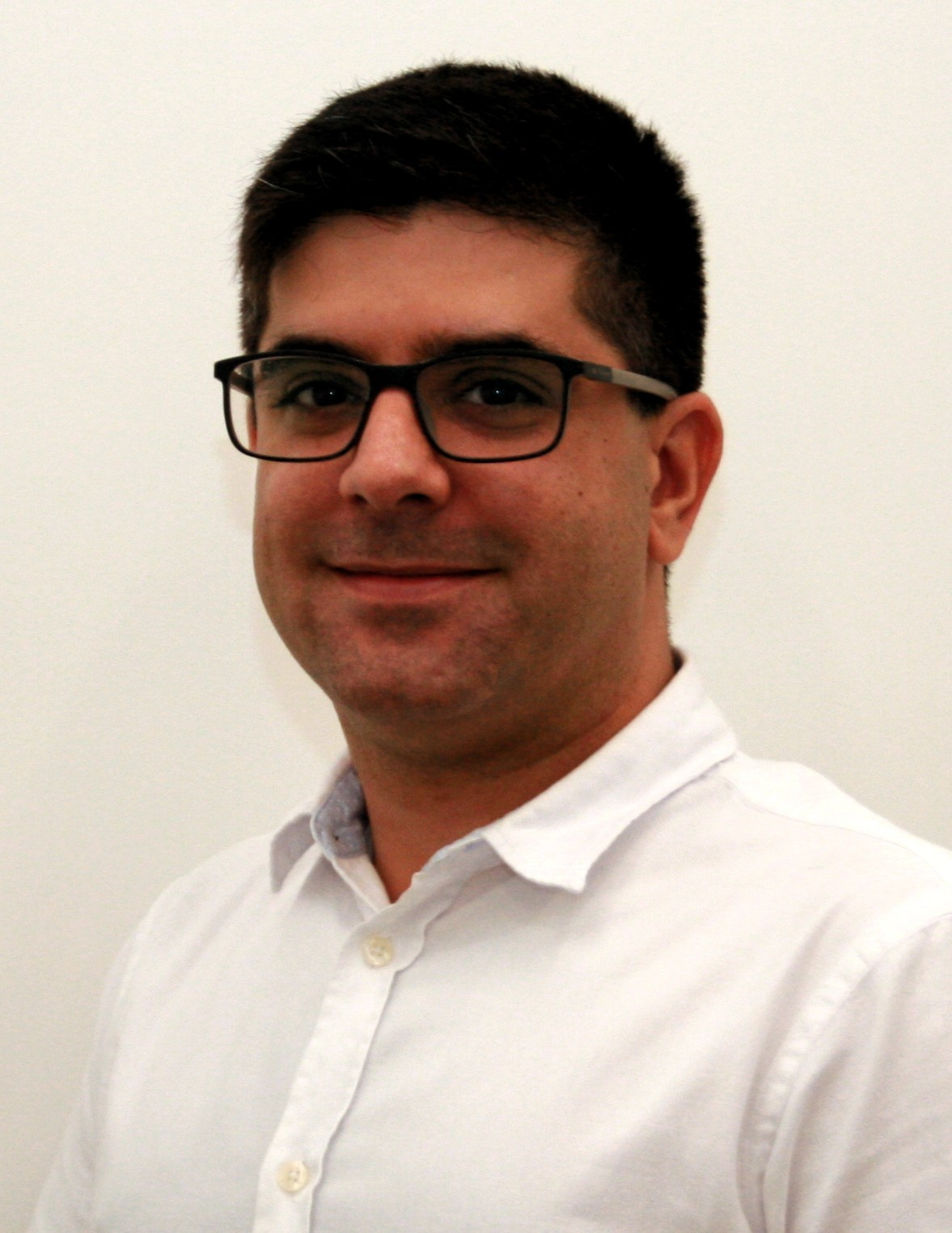}}]{Bruno C. Albertini}
Graduated in Computer Engineering from the Federal University of Rio Grande (2005), master's degree (2007) and PhD in Computer Science from the State University of Campinas (2012) in the area of Computer Architecture. He currently holds the position of Professor at the Department of Computer Engineering and Digital Systems (PCS) of the Polytechnic School of the University of São Paulo (EPUSP). Has experience in Computer Engineering, with emphasis on Hardware, working mainly on the following topics: hardware simulation, computational reflection, platform-based hardware design, computer architecture, wireless sensor networks, embedded systems, cryptohardware and AI in hardware. He joined LAA (Laboratory of Agricultural Automation) and BioComp in 2014, and has since applied his research to the environment, biodiversity and agriculture.
\end{IEEEbiography}

\begin{IEEEbiography}[{\includegraphics[width=1in,height=1.25in,clip,trim={8mm 2mm 1mm 1mm}]{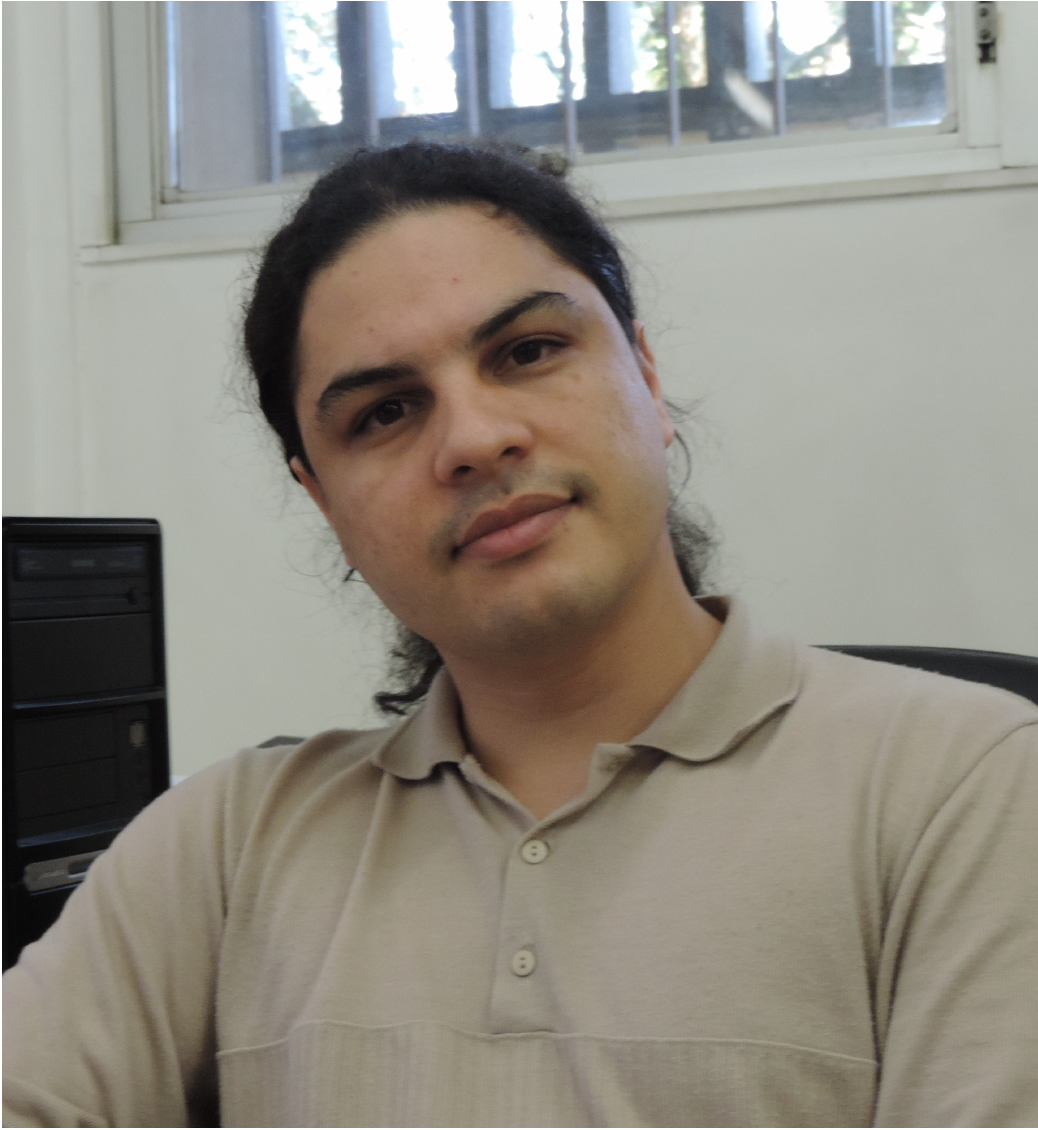}}]{Marcos A. Simplicio Jr.}
is an Associate Professor at Escola Politecnica, Universidade de Sao Paulo (USP). He has a Master degree (2006) in Engineering conferred by the Ecole Centrale des Arts et Manufactures (Ecole Centrale Paris), France, and received his Computer Engineering PhD from USP in 2010. His main research interests are cryptography, cybersecurity, and distributed systems. 
\end{IEEEbiography}

\end{document}